%% file: main.tex
\documentclass[conference]{IEEEtran}
\usepackage[utf8]{inputenc}   % Encoding of this TeX source file
\usepackage[T1]{fontenc}
\usepackage{csquotes}         % Proper quoting
\usepackage[nocompress]{cite} % Citation numbers automatically sorted and properly ranged, toggles: nocompress, noadjust
\usepackage{microtype}
\usepackage{balance}
\usepackage{array}
\usepackage{graphicx}
\usepackage{tikz}             % For copytight notice
\usepackage{amsmath,amssymb}
\usepackage[LGRgreek]{mathastext}  % Text in math mode
\usepackage{enumitem}

\usepackage{inconsolata}
\usepackage{listings}
\usepackage{xcolor}
\definecolor{logging-blue}{RGB}{1, 130, 172}
\usepackage{url}
\definecolor{dark-red}{rgb}{0.3,0.1,0.1}
\definecolor{dark-green}{rgb}{0.1,0.3,0.1}
\definecolor{dark-blue}{rgb}{0.1,0.1,0.5}
\usepackage[colorlinks,linkcolor=dark-red,citecolor=dark-green,urlcolor=logging-blue]{hyperref}
\usepackage[nameinlink]{cleveref}

\newcommand\copyrighttext{%
  \footnotesize \textcopyright 2021 IEEE. Personal use of this material is permitted. Permission from IEEE must be obtained for all other uses, in any current or future media, including reprinting/republishing this material for advertising or promotional purposes, creating new collective works, for resale or redistribution to servers or lists, or reuse of any copyrighted component of this work in other works. Cite this article as follows: V. Švábenský, J. Vykopal, D. Tovarňák, and P. Čeleda. \textit{Toolset for Collecting Shell Commands and Its Application in Hands-on Cybersecurity Training}, in Proceedings of the 51st IEEE Frontiers in Education Conference (FIE '21). Lincoln, Nebraska, USA, 2021. DOI: \href{https://doi.org/10.1109/FIE49875.2021.9637052}{10.1109/FIE49875.2021.9637052}.}
\newcommand\copyrightnotice{%
\begin{tikzpicture}[remember picture,overlay]
\node[anchor=south,yshift=12pt] at (current page.south) {\fbox{\parbox{\dimexpr\textwidth-\fboxsep-\fboxrule\relax}{\copyrighttext}}};
\end{tikzpicture}%
}

\newcommand{\circlednumber}[1]{\raisebox{.5pt}{\textcircled{\raisebox{-.9pt} {#1}}}}
\begin{document}

\title{Toolset for Collecting Shell Commands and Its Application in Hands-on Cybersecurity Training}

\author{\IEEEauthorblockN{Valdemar Švábenský}
\IEEEauthorblockA{Masaryk University\\
Brno, Czech Republic \\
svabensky@ics.muni.cz}
\and
\IEEEauthorblockN{Jan Vykopal}
\IEEEauthorblockA{Masaryk University\\
Brno, Czech Republic \\
vykopal@ics.muni.cz}
\and
\IEEEauthorblockN{Daniel Tovarňák}
\IEEEauthorblockA{Masaryk University\\
Brno, Czech Republic \\
tovarnak@ics.muni.cz}
\and
\IEEEauthorblockN{Pavel Čeleda}
\IEEEauthorblockA{Masaryk University\\
Brno, Czech Republic \\
celeda@ics.muni.cz}
}

\maketitle
\copyrightnotice

% 304 words
\begin{abstract}
This Full Paper in the Innovative Practice category presents and evaluates a technical innovation for hands-on classes.
When learning cybersecurity, operating systems, or networking, students perform practical tasks using a broad range of command-line tools. Collecting and analyzing data about the command usage can reveal valuable insights into how students progress and where they make mistakes. However, few learning environments support recording and inspecting command-line inputs, and setting up an efficient infrastructure for this purpose is challenging.
To aid engineering and computing educators, we share the design and implementation of an open-source toolset for logging commands that students execute on Linux machines. Compared to basic solutions, such as shell history files, the toolset's novelty and added value are threefold. First, its configuration is automated so that it can be easily used in classes on different topics. Second, it collects metadata about the command execution, such as a timestamp, hostname, and IP address. Third, all data are instantly forwarded to central storage in a unified, semi-structured format. This enables automated processing of the data, both in real-time and post hoc, to enhance the instructors' understanding of student actions. The toolset works independently of the teaching content, the training network's topology, or the number of students working in parallel.
We demonstrated the toolset’s value in two learning environments at four training sessions. Over two semesters, 50 students played educational cybersecurity games using a Linux command-line interface. Each training session lasted approximately two hours, during which we recorded 4439 shell commands. The semi-automated data analysis revealed different solution patterns, used tools, and misconceptions of students. Our insights from creating the toolset and applying it in teaching practice are relevant for instructors, researchers, and developers of learning environments. We provide the software and data resulting from this work so that others can use them in their hands-on classes.
\end{abstract}


% =============== Section start ===============
\section{Introduction}
\label{sec:intro}

% Cybersecurity must be practiced hands-on
Hands-on training is vital for gaining expertise in computing disciplines. Topics such as cybersecurity, operating systems, and networking must be practiced in a computer environment so that students can try various tools and techniques. Such a learning environment is called a \textit{sandbox}. It contains networked hosts that may be intentionally vulnerable to allow practicing cyber attacks and defense. These skills are grounded in the current cybersecurity curricular guidelines~\cite{cybered} to address the increasing shortage of cybersecurity workforce~\cite{isc2}.

% CLI is important
For the training, each student receives an isolated sandbox hosted locally or in a cloud. To solve the training tasks, students work with many tools, both in a graphical user interface (GUI) and a command-line interface (CLI). This paper focuses on the Linux CLI, which is common in higher education of computing, as well as software development in the industry practice.

% Research problem statement
Analyzing CLI interactions opens opportunities for educational research and classroom innovation. In traditional face-to-face classes, instructors must look at the students' computer screens to observe the learning process. However, this approach does not scale for large classes, and it becomes difficult for distance education. Instead, if the students' executed commands are logged, instructors and researchers may leverage them to support learning. By employing the methods of educational data mining~\cite{handbook-edm2010} and learning analytics~\cite{handbook-la2017}, the CLI data can help achieve important educational goals, such as to:
\begin{itemize}
    \item better understand students' approaches to learning, both in face-to-face and remote classes,
    \item objectively assess learning, and
    \item provide targeted instruction and feedback.
\end{itemize}

This paper examines the following research question relevant for instructors: \textit{What can we infer from students' command histories that is indicative of their learning processes?} Specifically, our goal is to understand how students solve cybersecurity assignments by analyzing their CLI usage. To address this question, we propose a generic method for collecting CLI logs from hands-on training. Then, we evaluate this method by gathering the logs from 50 students at four training sessions and investigating three sub-questions / use cases of the data:
\begin{enumerate}
    \item \textit{What does the command distribution indicate about the students' approach to solving the tasks?} Our motivation is to analyze which tools are commonly used and how effective they are with respect to the training tasks.
    \item \textit{Which commands are used immediately after the student accesses the learning environment?} We can observe if the students started solving the initial task, familiarized themselves with the environment, or displayed off-task behavior. This allows for providing suitable scaffolding.
    \item \textit{How much time do students spend on the tasks, and how often do they attempt an action?} Observing the time differences between successive commands can indicate the students' skill level and support assessment.
\end{enumerate}

% Contribution, novelty
Although there are many cybersecurity learning environments, which we review in \Cref{sec:related-work}, their logging support is often limited or non-existent. The current solutions do not allow instructors to uniformly collect CLI data and metadata with minimal setup and then correlate the logs from multiple sandboxes for advanced analyses. 

This paper addresses this gap by presenting and evaluating a technical innovation for hands-on classes that employ CLI tools. We created a toolset that collects Linux shell commands in physical or virtual learning environments and stores them in a unified format. Compared to the previous practice, where observing students' learning was difficult or even impossible, the proposed innovation enables understanding student approaches at scale. It also allows employing educational data mining and learning analytics techniques to gain further insights.

% Paper outline
The toolset design is explained in \Cref{sec:toolset}. In \Cref{sec:study}, we introduce a study that deploys the toolset in practice and evaluates it in authentic educational contexts. \Cref{sec:results} presents the results of the study and addresses the questions above. \Cref{sec:discussion} discusses the study and proposes multiple research ideas that further leverage the collected data. Finally, \Cref{sec:conclusions} summarizes our contributions. We also publish the toolset as open-source software. Instructors, researchers, and developers can use it to enhance computing classes, such as teaching cybersecurity, operating systems, and networking.

% =============== Section start ===============
\section{Related Work}
\label{sec:related-work}

This section contextualizes the work within the pedagogical literature. Then, it reviews learning environments and research in collecting and analyzing student data. We also point out the addressed gaps and explain how we differ from existing work.

\subsection{Theoretical and Pedagogical Background}

\textit{Active learning} is \enquote{any instructional method that engages students in the learning process}~\cite{Prince2004}. Sanders et al.~\cite{Sanders2017} reviewed 38 active instructional techniques, including hands-on labs, educational (serious) games, and automated tutoring systems that provide feedback to students. Unlike traditional lectures, active learning enables students to understand the topic more deeply~\cite{Petty2009}. Although the context matters, research strongly supports the notion that incorporating active methods improves learning~\cite{Prince2004}. That is why practice with CLI tools is important in engineering and computing education.

\textit{Zone of proximal development}~\cite{handbook-CER9} is a learning theory about an imaginary \enquote{space} slightly beyond the student's comfort zone, but where the student can learn if given some \textit{scaffolding}. Scaffolding is instructional support such as feedback and advice from the teacher, assignment of tasks suitable for the learner's skill level, and provision of hints. Applying this cognitive science theory improves the effectiveness of hands-on training. Although students often face ill-defined challenges that require complex problem-solving skills, given suitable scaffolding, they learn to solve practical problems in an authentic setting. Observing students' learning processes by employing CLI logging is the first step toward providing scaffolding and timely feedback, which we demonstrate in \Cref{subsec:futurework}.

\subsection{Learning Environments for Cybersecurity Training}
\label{subsec:related-work-env}

% Platform = A computer or hardware device and/or associated operating system, or a virtual environment, on which software can be installed or run.
% https://csrc.nist.gov/glossary/term/Platform

% Testbed = A platform for conducting rigorous and replicable testing of scientific theories, computational tools, and new technologies.
% https://en.wikipedia.org/wiki/Testbed
% https://www.nist.gov/labs-major-programs/research-test-beds
% https://www.mscoe.org/

% Testbeds and cyber ranges: definition, examples
A \textit{security testbed} is a physical or virtual environment that emulates computer systems for research, development, or education. It allows reliable, accurate, repeatable, and safe execution of experiments~\cite{siaterlis2014}. A related term is a \textit{cyber range}, which is a platform for interactive simulation environments~\cite{ecso2020} for security training and experimentation~\cite{davis2013}. Examples of testbeds and cyber ranges used in cybersecurity education are DETERLab~\cite{benzel2011, mirkovic2014}, EDURange~\cite{weiss2017magazine}, GENI~\cite{berman2014, mountrouidou2018}, and KYPO CRP~\cite{my-2021-FIE-kypo-csc}. For a thorough overview, see~\cite{yamin2020}.

% Virtual machines
Testbeds and cyber ranges may be complex and costly to operate. However, packaged virtual machines (VMs) have similar benefits and are simpler to use. Thus, they are widely employed in cybersecurity courses~\cite{oleary2017}. Authors of~\cite{Timchenko:2015} teach security with locally hosted VMs. Projects SecKnitKit~\cite{siraj2015} and SEED~\cite{du2011} provide VMs with labs for guided teaching of security topics. Some SEED labs were ported into Labtainers, a training framework based on Docker containers to allow easier use~\cite{Irvine2017}. The containers, similarly to VMs, ensure that the lab software will always run the same. Finally, Vulnhub~\cite{vulnhub} and Metasploitable~\cite{metasploit} share virtual targets for hacking. 

% Comment on logging in testbeds, cyber ranges, and VMs
Although many learning environments exist, few of them leverage the potential of collecting and analyzing training data. Only EDURange, DETERLab, and Labtainers include host-based logging of commands (see \Cref{subsec:related-work-data} for comparison with our approach). The VM solutions do not provide an easy way to monitor student actions~\cite{weiss2017magazine, weiss2016}. What is more, creating and configuring VMs is time-consuming, and they remain static once created~\cite{schreuders2017}.

\subsection{Collection and~Analysis of Cybersecurity Education Data}
\label{subsec:related-work-data}

% Related research
In Capture the Flag (CTF) challenges, the participants solve cybersecurity tasks to obtain text strings called \enquote{flags} as proof of solution. Flags include, for example, passwords from a breached service. Previous works collected quantitative data from CTF, such as the flag content or timing~\cite{oslejsek2019}. Although such data are relevant for assessing success or failure, they are insufficient for understanding \textit{how} students learn. To illustrate, an incorrect flag means that the student did not complete the challenge. However, it does not reveal the approach to obtaining the flag, making it harder to address the student's misconceptions. We need detailed information about how the student progressed while solving the tasks. Within the scope of our work, this means the entered shell commands.

% Keyloggers
Keyloggers such as \texttt{ttylog}~\cite{ttylog} or recording software such as \texttt{asciinema}~\cite{asciinema, Hassan2020} appear to be a universal solution. A keystroke logging approach has also been used in programming education research~\cite{ihantola2015}. However, keyloggers capture individual characters, which is too fine-grained. Reconstructing the commands then requires parsing large amounts of data. Video recording software, on the other hand, does not allow easy correlation of data from multiple sandboxes. Native Linux solutions, such as commands stored in the \texttt{.bash\_history} file at a local host, have the same limitation. Therefore, we need a compact machine-readable format suitable for logs from cybersecurity training.

% Mitigation: command-line history
Irvine et al.~\cite{Irvine2017} collected files containing the \texttt{stdin} and \texttt{stdout} streams from Docker containers of students working in Labtainers. Labtainers software then used the files to check if a student achieved a certain goal (such as completing a learning task). However, the data were not aggregated in real-time from all containers to gain an overall understanding. Students had to send the log files manually after the training.

Weiss et al.~\cite{weiss2017magazine, weiss2016} logged shell commands in EDURange to visualize the students' steps. In follow-up work, Mirkovic et al.~\cite{mirkovic2020} presented the system for logging and analyzing commands in DETERLab. The system monitored terminal input and output and compared them with task milestones to automatically assess student progress. Similarly~\cite{andreolini2019framework, tian2018real} also implemented host-based logging to reconstruct the students' progress. Although the goal of the above-mentioned works is similar to ours, their approach is complementary. We extend the state of the art in the following aspects:

\begin{enumerate}
    \item Focus on understanding of students' learning processes, not only the outcomes (succeeding or failing a task) when analyzing data from cybersecurity training (see \Cref{sec:results}).
    \item Real-time, automatic, and secure forwarding of logs.
    \item Support of more Linux-based operating systems.
    \item Enabling the logging of commands from Metasploit shell~\cite{metasploit}, an essential cybersecurity tool.
    \item Easier deployment to other existing learning environments. The deployment is automated using Ansible~\cite{ansible}, a well-established software for configuration management.
    \item Generic output format that can be processed by custom software or integrated with an existing one, such as Weka~\cite{weka} or ELK~\cite{elkstack}.
    \item Demonstration of innovative analyses of command histories (see proof of concept in \Cref{subsec:futurework}).
\end{enumerate}

\subsection{Tools for Teaching Linux Shell Commands}

To provide a complementary viewpoint, we briefly examine systems for teaching Linux CLI in general, without an explicit focus on cybersecurity. TuxLab~\cite{tuxlab} and uAssign~\cite{uassign2019} are two such systems. Both leverage virtualization, since they are based on Docker containers. Students can use them for practicing Linux commands, which are immediately evaluated in the container and autograded. The difference from our work is that neither of them collects the commands for further analysis. TermAdventure~\cite{Suppa2021}, on the other hand, stores the command logs, but is not a standalone logging toolset as we present here.

% =============== Section start ===============
\section{Toolset for Command Logging}
\label{sec:toolset}

\begin{figure*}[t]
\centering
\includegraphics[width=\linewidth]{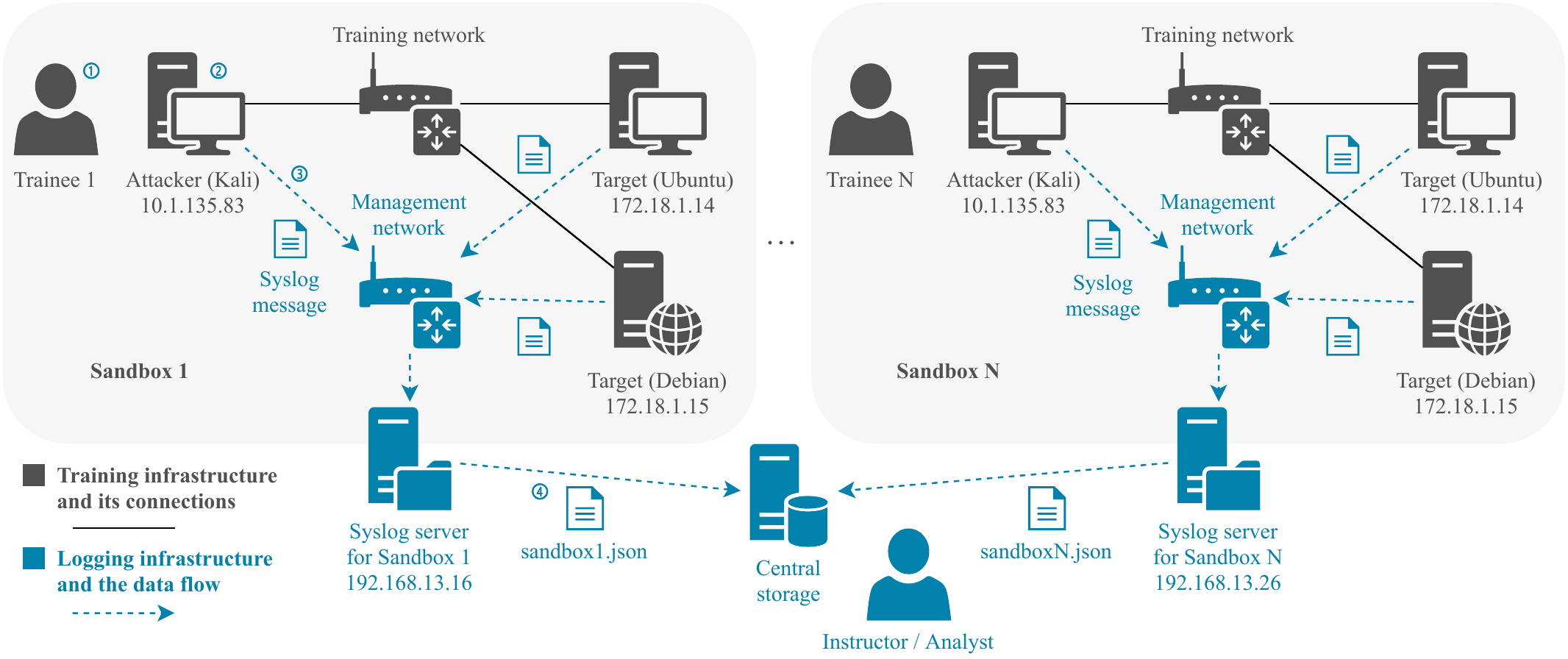}
\caption{Toolset design and data flow. \circlednumber{1} The trainee executes a command on the Attacker machine. \circlednumber{2} The text log in \Cref{fig:command-local} is generated and stored on the local machine. \circlednumber{3} Logs from all machines in a sandbox are forwarded to the Syslog server. \circlednumber{4} Logs from Syslog servers of all sandboxes are forwarded to the central storage and formatted as JSON in \Cref{fig:command-json}, which instructors can access and analyze.}
\label{fig:overview}
\end{figure*}

\input{figures/command2.tex}
\input{figures/command3.tex}

\Cref{fig:overview} shows the generic topology of a sandboxed network for cybersecurity training. It contains an attacker machine controlled by the trainee and several vulnerable hosts (black-colored elements). Since we assume one trainee per sandbox,\footnote{If teamwork is needed, a group can access the same sandbox. However, the toolset cannot distinguish individuals within the group. The mapping of logs to students is done only based on the sandbox ID.} each trainee receives its identical copy. On top of that, we configure the (blue-colored) infrastructure elements responsible for collecting CLI logs from the hosts.

\subsection{How Does the Toolset Work?}

Each host in the sandbox has two network interfaces: \textit{training} (black) and \textit{management} (blue). The former enables the trainee to interact with hosts in the sandbox and complete the training tasks. The latter exists in the network to initially configure the sandbox. We utilize it for host-based logging and forwarding of the logs to central storage.

% Pipeline of the data collection
The logs are generated on all sandboxed machines that a trainee accesses. Every time (s)he enters a command on a Linux host, it is stored on that host using the Syslog protocol~\cite{rfc5424}. The log is then instantly forwarded to the Syslog server of the sandbox. Finally, the logs from Syslog servers of all sandboxes are securely forwarded to a central storage (see \Cref{fig:overview}). The intermediate step with the Syslog server is optional and can be bypassed to log directly into the central storage. The data can be accessed continuously in real-time or after the training.

To verify that all submitted commands arrive correctly, we performed multiple manual tests. We deployed the data collection in two learning environments~\cite{my-2021-FIE-kypo-csc}: a complex cloud-based platform KYPO CRP and a lightweight, locally virtualized platform Cyber Sandbox Creator. \Cref{sec:study} and \Cref{sec:results} present the study that analyzes the collected data.

The configuration files that supplement this paper (see \Cref{subsec:contributions}) enable others to deploy this toolset in their physical or virtual environment. Instructors who do not have access to a cloud can prepare VMs provisioned with the logging. Students can download them and use them on their computers, and instructors can observe the commands at the central storage deployed at their institution or in a public cloud.

\subsection{Implementation Details}

Our approach works for the Bash shell~\cite{FSFBash2019} since it employs a Bash variable \texttt{\$PS0}. Its value is expanded \enquote{after reading a command and before the command is executed}~\cite{FSFBash2019} to a call of the \texttt{logger}~\cite{logger} command, which can enter arbitrary messages into the host's system log socket (typically \texttt{/dev/log}). This \texttt{AF\_LOCAL} socket is widely used by standard logging daemons and services. Regardless of the used Linux distribution, modern logging daemons can transform, store, and forward log entries flexibly. We use \texttt{rsyslogd} daemon~\cite{rsyslog} to forward the logs to a remote location via the Syslog protocol, as described above. Since the logging is based on Syslog, it can be deployed generically. Any application can log into Syslog, or at least in a plain text file that can be forwarded to Syslog.

\subsection{What Data Are Collected?}

\Cref{fig:command-local} shows a command executed on the Attacker machine and the subsequently generated log record. We collect each line of the terminal input: the entire command, including all arguments. All inputs are timestamped (up to microsecond precision) and identify the host and sandbox in which the command was executed. The hosts must have a synchronized time to allow correlating the commands. We also collect essential metadata about the execution, such as the working directory and type of the command to distinguish between the Bash and Metasploit shell.

To protect the trainees' privacy, no personally identifiable information is collected. The sandbox is identical for all trainees, so the usernames are generic and unrelated to their identity. Since the sandbox is isolated, data about the trainees' physical computers are not logged or even accessed.

The captured data are stored in JSON shown in \Cref{fig:command-json}. This universal, machine-readable format allows data analysts to process the logs in a way that suits their needs. This can involve writing custom scripts, using open-source tools such as Weka~\cite{weka}, or adding ELK~\cite{elkstack} to the analysis pipeline.

\subsection{Benefits of the Logging Toolset}

The logging toolset has the following properties (some are based on the ISO/IEC SQuaRE standard~\cite{SQUARE2011}):
\begin{itemize}
    \item \textit{Flexibility}: it can be added to or removed from an arbitrary Linux host, regardless of the content of the training.
    \item \textit{Automated deployment}: it is automatically applicable to any number of sandboxes. This saves the instructors' time and allows them to monitor multi-stage attack scenarios.
    \item \textit{User transparency}: the logging interface is in the isolated management network, so it does not affect the training network or negatively impacts trainees' learning or user experience.
    \item \textit{Live streaming}: each new log record is immediately forwarded to the central storage. Instructors can analyze the data instantly and provide rapid feedback to trainees.
    \item \textit{Remote storage}: even if the logs are modified locally, the copy on the central storage stays intact. If the connection is interrupted, all logs are buffered in the local machine's memory and then forwarded once it is back online.
    \item \textit{Secure transfer of data}: the logs are signed and encrypted to maintain their integrity and confidentiality.
    \item \textit{Machine-readable output}: the logs are stored in a universal, semi-structured format (JSON), which simplifies their further processing.
    \item \textit{Installability}: the toolset is open-source and uses only standard Linux components. Thus, it is widely applicable and compatible.
\end{itemize}

\subsection{Limitations of the Logging Toolset}
\label{subsec:limitations}

The practical application of the toolset in two learning environments revealed some of its shortcomings:
\begin{itemize}
    \item The data collection is constrained to the commands entered in a shell. Any edits of commands before the submission (pressing the Enter key) are not recorded. We also do not capture any other data indicating progress, such as mouse clicks, network traffic, or filesystem changes.
    \item We do not collect the command's return code or output, since \texttt{logger} is called before the command is executed. Neither we collect the submission of control sequences such as Ctrl + C or responses to prompts.
    \item The logging currently works only on Linux hosts. However, it could be analogously extended to Windows hosts to log PowerShell~\cite{Powershell} instead of Linux shell commands.
    \item An advanced trainee with administrator privileges on a machine could tamper with or turn off the logging from that host. However, the instructor would notice the absence of logs. A future challenge is to hide the logging to prevent both intentional and accidental changes.
\end{itemize}

% =============== Section start ===============
\section{Study Setup and Methods}
\label{sec:study}

We now describe the study that involved deploying our toolset in authentic educational settings on multiple occasions. The study addresses our goal of understanding trainees' learning processes by examining their command histories. Moreover, we evaluate the applicability of the toolset and demonstrate the value of the collected data.

\subsection{Teaching Context and Training Content}

During the Fall 2019 and Spring 2020 semesters, the students of an undergraduate course Seminar on the Simulation of Cyber Attacks~\cite{svabensky2018kypolab} at Masaryk University created eight educational cybersecurity games. The games involve practicing security skills using Linux Bash shell in a virtual network similar to the one in~\Cref{fig:overview}. The game tasks focus on gradually compromising the security of a simulated IT system, following the stages of Mandiant’s Attack Lifecycle~\cite[p. 27]{attacklifecycle}. All the games exploit the same vulnerabilities, and they differ only in the fictitious background story and the task presentation.

The games' learning objectives and tasks are centered around using CLI tools in Linux distribution called Kali~\cite{kalitools, ahmad2019}. Each game starts by scanning a target network using \texttt{nmap}~\cite{nmap}. Nmap can determine what services are running on the network hosts and their port numbers. This corresponds to the first stage \textit{Initial Reconnaissance} in the Mandiant’s Attack Lifecycle. The next task is to identify a vulnerable service on the target and exploit it using Metasploit~\cite{metasploit}, a penetration testing software that contains exploits for various vulnerabilities. After exploiting the service, the attacker accesses the machine, copies a private SSH key, cracks its passphrase, and uses the key to connect to another host via SSH. For the passphrase cracking, John the Ripper tool was recommended. Given a text file with passphrases, the \texttt{john} command performs a dictionary attack.

When the students of the course finished creating their games, we added the logging module into the game configuration files. Although the games differed in technical details, the toolset was applicable to all of them. Afterward, all games were deployed in the KYPO CRP~\cite{my-2021-FIE-kypo-csc}.

\subsection{Participants}

From December 2019 to July 2020, we hosted four events (game sessions), during which a total of 50 participants (\textit{trainees}) played one of the eight games mentioned above. Each event lasted approximately two hours and had a diverse participant population:
\begin{itemize}
    \item undergraduate or graduate students of computer science at our university, who have a basic cybersecurity background,
    \item security professionals, and
    \item senior high school students and bachelor students of other universities, the finalists of the Czech national cybersecurity competition.
\end{itemize}

The trainees attended voluntarily because of their interest in security and were not incentivized in any way. None of the trainees played any of the eight games before.

\subsection{Privacy and Ethical Measures}

Before the training sessions, we discussed the research with the institutional review board of our university. We obtained a waiver from the ethical committee since we intentionally do not collect any personal information; the data are anonymous and cannot be linked to individuals. The trainees agreed to the anonymized data collection via informed consent. We ensured they would not be negatively affected by the research, and they had the right to stop participating at any time without any restrictions. After the data collection, we manually checked whether they typed any personal information in the CLI and anonymized it. Only one student typed his real university login.

\subsection{Methods of Data Processing and Analysis}

Since all games had identical learning objectives and vulnerabilities, we merged all the logs. Then, we wrote Python scripts that parsed and analyzed the data. In addition to quantitative statistics, two security instructors (authors of this paper) exploratively analyzed the output of the automated data processing. Their goal was to manually uncover deeper insights. Since the scripts automatically presented aggregated results, their interpretation was straightforward.

The analysis focused on cybersecurity tools. When preprocessing the data, we removed extraneous whitespace and, when applicable, disregarded the order of the arguments, which was sometimes irrelevant. Apart from this sanitation, we did not modify the trainee data.

% =============== Section start ===============
\section{Results}
\label{sec:results}

This section presents the analysis results to answer the three questions posed in \Cref{sec:intro}. Each question is addressed in a separate subsection. The results demonstrate the value of the collected command-line data. We also report our teaching experience, educational insights, and implications for practice. 

\subsection{Use Case 1: Executed Commands}

\Cref{tab:desc} provides an overview of the collected commands. The most frequently used cybersecurity tool was \texttt{nmap} with 240 occurrences out of the 4439 commands in the whole dataset. This was expected, since the initial task in all games was to scan the target network. When looking not only at cybersecurity-specific commands, \texttt{ls} was the most frequent (568 times), followed by \texttt{cd} (320 times), which corresponds to the nature of the training that required traversing the filesystem.

\input{tables/1-desc-stats.tex}

We now analyze how the trainees used three key command-line tools in the games: \texttt{nmap}, \texttt{ssh}, and \texttt{john}. We also examine the logs from the Metasploit console.

\subsubsection{Terminal Commands}

Nmap requires a target specification, such as a hostname or an IP address. However, in 14 cases, the trainees attempted an \texttt{nmap} scan but did not specify a target, which indicates misconceptions. Another interesting observation was that although the task was to scan ports, the option \texttt{-sn} that \textit{disables} the port scan was used 13 times. An IP address that belonged to a router was scanned 7 times. In 8 cases, the trainee made a typo in the target IP address since it was off by one digit, but 9 other times they wanted to scan an address that was not even in the training sandbox.

Using these findings, instructors can provide targeted feedback to students, for example, explain the causes of the mistakes to the class. They can also use these insights to devise a scoring rubric for assessment (for example, scanning a correct IP address: 1 point).

Later in the game, the trainees obtained a private SSH key. They had to crack its passphrase by executing a dictionary attack using \texttt{john}. The most common mistakes among 201 \texttt{john} commands in the dataset were not specifying the dictionary or setting the option \texttt{-}\texttt{-}\texttt{wordlist} to a folder instead of a file. After cracking the passphrase, the trainees usually had no problem connecting to a target host using \texttt{ssh}.

\subsubsection{Metasploit Commands}

Another task was to exploit a host using Metasploit. The correct course of action was:
\begin{enumerate}
    \item find suitable exploit modules using \texttt{search},
    \item select a correct module from the list using \texttt{use},
    \item configure module options using \texttt{set}, and
    \item run the module using \texttt{exploit} or \texttt{run}.
\end{enumerate}

The logs revealed the following misconceptions and unusual activities of the trainees:
\begin{itemize}
    \item instantly selecting the correct module without searching for it in Metasploit, which suggests previous knowledge of the attack or outside help,
    \item selecting a wrong exploit and attempting to run it,
    \item attempting to configure the module parameters with \texttt{set} before selecting the module with \texttt{use},
    \item incorrectly setting the machine initiating the exploit (parameter \texttt{LHOST}) to the target host or other values,
    \item attempting to run the exploit before setting its parameters.
\end{itemize}

\subsection{Use Case 2: First Commands Entered}

We reconstructed the first high-level action of each trainee from the first few commands. To illustrate, \Cref{tab:timeline} is an example of the first few commands of one trainee who started scanning the target network. The trainee’s intention was apparent despite some errors and irrelevant commands.

\input{tables/2-timeline}

This work pattern was also the most commonly observed. For 30 out of the 50 trainees, their first command was \texttt{nmap}, in which half of them correctly set a target. In five other cases, trainees started by familiarizing themselves with the network configuration using \texttt{ifconfig} or \texttt{ping}.

At the start of the training, the trainees did not deviate from the task and did not execute unusual commands. Nevertheless, analyzing the first actions can still be beneficial. If an instructor sees that a trainee started correctly, (s)he may devote more time to aid others. If an instructor observes off-task behavior at the beginning, (s)he can intervene early and help the trainee by providing targeted hints.

\subsection{Use Case 3: Time-based Statistics}

\Cref{tab:rq3} shows the statistics for two variables. The first is the time spent playing, calculated as the time difference between the first and the last command. The trainees usually spent about an hour on the training, which is enough to record interactions in depth. A slight limitation is that we cannot determine the exact end time of the training from the command-line logs only. The last submitted command does not necessarily mean the end of the work on the task. The student could have continued by using a GUI tool, consulting their progress with another person, or searching for online help, for example.

\input{tables/3-timings}

As expected, the time spent in the game positively correlates with the number of submitted commands (Spearman's $\rho = 0.79$). This correlation is statistically significant (two-sided $p$-value $< 10^{-11}$, calculated by the Python \texttt{scipy.stats} module~\cite{scipy}).

The second variable in \Cref{tab:rq3} is the time difference between two successive commands. The median, average, and standard deviation were first computed individually for each trainee. The table reports the averages of these individual quantities. The trainees usually submitted a command every minute or so, which seems appropriate since they also read the documentation and contemplated their approach.

The near-zero time differences occurred when a trainee copy-and-pasted a command or used shortcuts such as \enquote{arrow up} to repeat the previous command. Minor time differences were observed when a trainee made an error, immediately recognized it, and fixed it (see the last two lines in \Cref{tab:timeline}). We also observed differences smaller than the 15-second median when the trainee was unsure how to use a command and applied a trial-and-error method. Finally, a burst of commands in a short period may indicate scripting or attempts to tamper with the system, although we did not observe this in our data. On the other hand, long idle periods can mean disengagement, need for help, or technical issues with the learning environment.

% =============== Section start ===============
\section{Study Discussion}
\label{sec:discussion}

This section discusses the study limitations and proposes future work that will use the command-line data.

\subsection{Limitations of the Study}

The results have two main limitations. First, the logs can only indicate a problem, not explain why it occurs. We proposed possible explanations, but they are yet to be confirmed in an in-depth qualitative study with students. Nevertheless, the logs authentically record the trainees' progress, creating a basis for targeted educational interventions.

Second, we focused only on the teacher's perspective and enhancing the instructor's awareness of the classroom. It would be interesting to close the feedback loop and evaluate how this educational intervention impacts the students.

\subsection{Educational Implications and Future Work}
\label{subsec:futurework}

A CLI is difficult for beginners since it lacks a graphical interface and is strict about using the commands properly. Our toolset for collecting and analyzing the commands allows instructors to better understand students' behaviors, which is a prerequisite for helping them learn.

This paper demonstrated three initial use cases. For the future, we see more advanced research opportunities that require high-quality command-line data. The possibilities are to:
\begin{itemize}
\item Qualitatively analyze the causes of student mistakes and misconceptions (for example, using interviews and think-aloud protocols) and address them in the classroom.
\item Perform pattern mining to find interesting association rules and sequences in the student data and visualize them.
\item Provide automated, personalized feedback to many students at scale, both in real-time and after the training.
\item Assess students, including partial credit for unfinished solutions. Students can be graded for using (or not using) a specific set of commands and their arguments.
\item Measure learning in a pre-test/post-test study framework to see whether the interventions achieved by analyzing command-line data helped improve student achievement.
\end{itemize}

\begin{figure}[!ht]
\centering
\includegraphics[width=\linewidth]{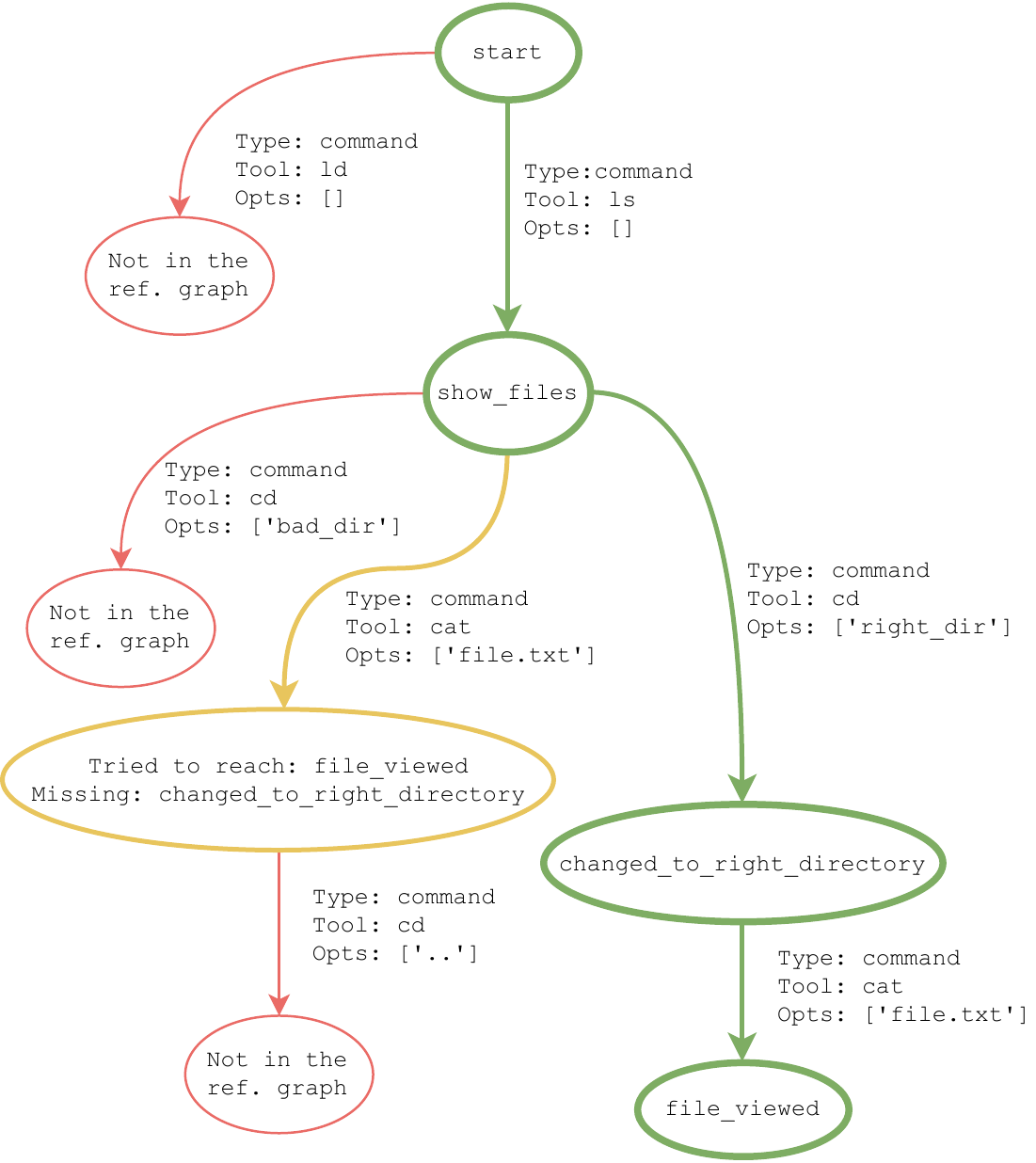}
\caption{Graph visualization of student's progress through an exercise that employs command-line tools.}
\label{fig:trainee-graph}
\end{figure}

\begin{figure}[!ht]
\centering
\includegraphics[width=\linewidth]{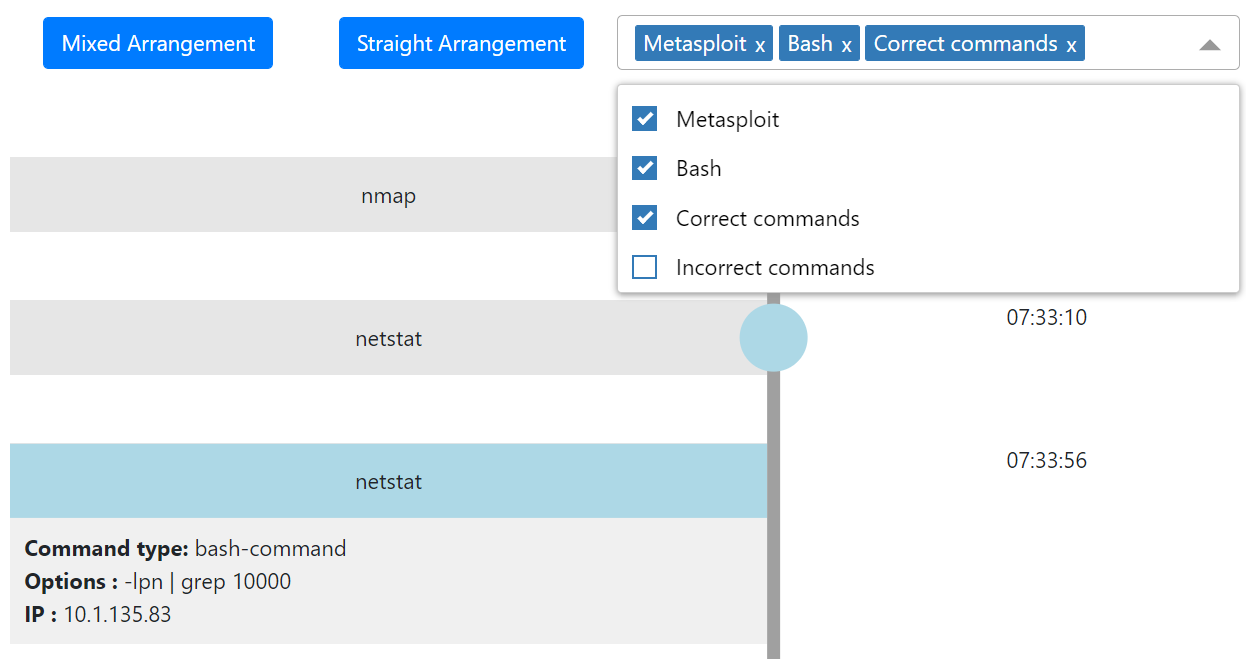}
\caption{Timeline visualization of student's progress through an exercise that employs command-line tools.}
\label{fig:timeline}
\end{figure}

In our further ongoing research, we investigate methods for graphical modeling of student's steps (see \Cref{fig:trainee-graph}). The proof-of-concept application automatically processes students' command histories and maps them to a reference solution of the training. It can display correctly achieved steps, errors, and potential omissions. Another proof-of-concept web application in \Cref{fig:timeline} displays an interactive timeline of the submitted commands, allowing to filter correct or erroneous attempts.

% =============== Section start ===============
\section{Conclusions}
\label{sec:conclusions}

Command-line tools are used in almost all areas of computing. In cybersecurity, they are essential both to learn about cyber attacks and to set up defense mechanisms. Even today, when security experts may also use sophisticated graphical interfaces, tools controlled via a shell still form an essence of cybersecurity operations. Therefore, learning and understanding these tools is crucial to mitigate the current cyber threats.

We presented a modular toolset for logging commands from Bash and Metasploit shell. It enables a data-driven understanding of students' approaches to practicing cybersecurity, system administration, and networking. We deployed the toolset in two learning environments and showed the value of the gained data by analyzing student approaches. Since the toolset can be applied in many other contexts, it can foster further research and development, as we showed in \Cref{subsec:futurework}.

Observing student actions in the training session is invaluable for instructors to provide timely help to students during the session or detailed feedback afterward. This is a substantial innovation of typical practice when no information or only the (in)correctness of the task solutions are recorded. Therefore, this paper contributes to addressing a challenge that computing instructors face. Although tinkering with a CLI is common for students who are learning, if errors occur repeatedly, instructors should know about that to intervene appropriately.

The logging toolset helps not only instructors and students of cybersecurity, but also in other courses that feature command-line tools, such as operating systems or networking. Observing student actions is especially relevant in distance education when instructors have limited awareness of what students do on their computers. This proved invaluable in our online teaching practice during the COVID-19 pandemic in 2020--2021. Most of the analyzed data were collected from remotely held training sessions, in which we would otherwise not have any awareness about the students' progress.

\subsection{Contributed Artifacts}
\label{subsec:contributions}

This paper's supplementary materials contain four core contributions for instructors, educational researchers, and developers of interactive learning environments.
\begin{enumerate}
    \item The toolset that can be integrated with existing learning environments. We share configuration scripts (Ansible~\cite{ansible} roles) that automatically deploy host-based logging to the training sandbox similar to the one in \Cref{fig:overview}.
    \item Educators can familiarize themselves with the toolset. They can automatically set up a learning environment from scratch on a personal computer using a single command. We enable this by providing a Vagrant~\cite{vagrant} definition file that builds a locally virtualized learning environment for a cybersecurity game. Moreover, the game includes exercise assignments that can be completed in the learning environment to practice cybersecurity skills.
    \item We publish the collected data from the training sessions studied in this paper, as well as others we conducted.
    \item We provide analytical software that was used to process and analyze the data.
\end{enumerate}

All artifacts are linked from a single Zenodo repository~\cite{dataset} and are available under permissive open-source licenses. Each artifact contains detailed instructions to allow educators, researchers, and developers to use our results in their contexts.

\section*{Acknowledgment}
This research was supported by ERDF project ``CyberSecurity, CyberCrime and Critical Information Infrastructures Center of Excellence'' (No. CZ.02.1.01/0.0/0.0/16\_019/0000822).

\balance
\bibliographystyle{IEEEtran}
\bibliography{references}

\end{document}

%% file: figures/command2.tex
\newcommand{\cmdpart}[2]{\underbrace{\texttt{#1}}_{#2}\ \ \ }

\begin{figure*}[!ht]
\begin{center}
\vspace*{-3mm}
\hrule
\vspace*{5mm}

\hspace*{-2.5mm}\circlednumber{1} Command executed by the trainee on the Attacker machine:\quad \verb!root@attacker:/home\# nmap --help!
\end{center}

\[
\setlength{\jot}{9pt} % Line spacing
\begin{split}
&\circlednumber{2}\ Log\hspace*{-1mm}:\ \cmdpart{Jul 03 2020 8:09:25}{timestamp}
\cmdpart{username=\textquotedbl root\textquotedbl}{username}
\cmdpart{attacker}{hostname}
\cmdpart{src=\textquotedbl 10.1.135.83\textquotedbl}{host\ IP\ address} \\
&\hspace*{12mm}
\cmdpart{wd=\textquotedbl /home\textquotedbl}{working\ directory}
\cmdpart{cmd=\textquotedbl nmap -\,-help\textquotedbl}{command}
\cmdpart{cmd\_type=\textquotedbl bash-command\textquotedbl}{command\ type}
\cmdpart{sid=\textquotedbl 1\textquotedbl}{sandbox\ ID}
\end{split}
\]
\caption{\circlednumber{1} A command executed on the Attacker machine in Sandbox 1 and \circlednumber{2} a corresponding log record stored locally.}
\label{fig:command-local}
\end{figure*}

%% file: figures/command3.tex
\begin{figure}[!ht]
\vspace*{-4mm}
\begin{lstlisting}[basicstyle=\small\ttfamily]
{
  "timestamp"  : "2020-07-03T08:09:25+01:00",
  "username"   : "root",
  "hostname"   : "attacker",
  "ip"         : "10.1.135.83",
  "wd"         : "/home",
  "cmd"        : "nmap --help",
  "cmd_type"   : "bash-command",
  "sandbox_id" : "1"
}
\end{lstlisting}
\caption{Content of the file \texttt{sandbox1.json} generated from the Syslog record in \Cref{fig:command-local} and saved in the central storage.}
\label{fig:command-json}
\vspace*{-3mm}
\end{figure}

%% file: tables/1-desc-stats.tex
\begin{table}[!ht]
\caption{Overall and per-trainee statistics of commands entered by~the~50 trainees (players of the cybersecurity games).}
\label{tab:desc}
\centering
\begin{tabular}{lrrrrrr}
\hline
Command type & Total & Min & Max & Median & Avg & Stdev \\
\hline
\textbf{Linux Bash} & 3121 & 4 & 266 & 45 & 62.4 & 57.3 \\
\textbf{Metasploit} & 1318 & 0 & 213 & 17 & 26.4 & 36.3 \\
\textbf{Both types} & 4439 & 5 & 358 & 68 & 88.8 & 74.9 \\
\hline
\end{tabular}
\end{table}

%% file: tables/2-timeline.tex
\begin{table}[!ht]
\caption{Timeline of the training's start for a selected trainee (shortened), along with a time delay between two commands.}
\label{tab:timeline}
\centering
\begin{tabular}{rrl}
\hline
Time    & Command & Command \\[-0.5mm]
{[m:ss]} & delay [s]      & \\
\hline
0:00 & N/A & \verb!nmap --help!\\
0:55 &  55 & \verb!nmap 172.18.1.5!\\
3:11 & 136 & \verb!pwd!\\
3:14 &   3 & \verb!ls!\\
3:26 &  12 & \verb!nmap --help!\\
4:10 &  44 & \verb!nmap -sV --p 10000 172.18.1.5!\\
4:17 &   7 & \verb!nmap -sV -p 10000 172.18.1.5!\\
\hline
\end{tabular}
\end{table}

%% file: tables/3-timings.tex
\begin{table}[!ht]
\caption{Total time spent on the training and the time differences (gaps) between two successive commands for the 50 trainees.}
\label{tab:rq3}
\centering
\begin{tabular}{lrrrrr}
\hline
 & Min & Max & Median & Avg & Stdev \\
\hline
\textbf{Game time [m:ss]} & 16:10 & 1169:20 & 77:12 & 108:12 & 159:05 \\
\textbf{Avg gap   [m:ss]} &  0:00 &  631:57 &  0:15 &   1:13 &  10:17 \\
\hline
\end{tabular}
\end{table}